\documentclass[preprint,aps]{revtex4-2}
\usepackage{color}
\usepackage{mathtools}
\usepackage{array}
\usepackage{tabularx}
\usepackage{diagbox}
\usepackage{tikz}
\usetikzlibrary{decorations.pathmorphing}
\usepackage{bm}
\usepackage{amsmath,bm,amssymb,amsfonts,dcolumn,color,graphicx,latexsym,epsfig}
\usepackage{rsfso}
\usepackage{hyperref}
\hypersetup{
colorlinks=true,
linkcolor=magenta,
filecolor=magenta,      
urlcolor=blue,
citecolor=blue,
}
\usepackage{multirow}
\usepackage{natbib}
\usepackage{float}
\usepackage{booktabs}
\usepackage{pifont}
\usepackage{mathrsfs}
\usepackage{caption}
\usepackage{caption, threeparttable}
\begin{document}

\title{Closed strings in a class of pp-wave spacetimes and the memory effect}
\author{Ayan Dey and Sayan Kar}
\email{ayan11@kgpian.iitkgp.ac.in, sayan@phy.iitkgp.ac.in}
\affiliation{Department of Physics, Indian Institute of Technology, Kharagpur 721 302, India}

\begin{abstract}
We study closed string evolution in the pp-wave spacetime assuming different
pulse shapes (square and sech-squared) in the exact gravitational wave metric. 
It is shown that
the shape of a circular closed string deforms permanently,
after the pulse has departed. The worldsheet geometry also displays characteristic permanent changes caused by the gravitational wave pulse. The above effects 
collectively demonstrate features 
akin to what is known as `memory' in gravitational wave physics. 
\noindent 
\end{abstract}


\maketitle

\newpage

\section{Introduction}
\noindent In studies on the physics of gravitational waves, the gravitational wave memory effect
is often illustrated by considering the change of shape of a ring of particles when it
encounters a gravitational wave pulse for a short duration of time \cite{favata1, favata2}. For example, if 
the circular shape of the ring of particles gets {\em permanently} deformed to an
elliptical (or any other) shape, we say that the pulse has `memory'. On the other hand, if the deformation is
not permanent and the ring reverts back to its circular shape we say that the pulse has
`no memory'. Such a visually appealing picture is very helpful in providing an elegant way
of appreciating what `memory' really means (see Figure 1). This illustrative example is by no means
a toy example--in fact it has in it the very basic idea of `permanent change' which
lies at the heart of all ways of defining memory.
\begin{figure}[h]
         \includegraphics[width=1.1\textwidth]{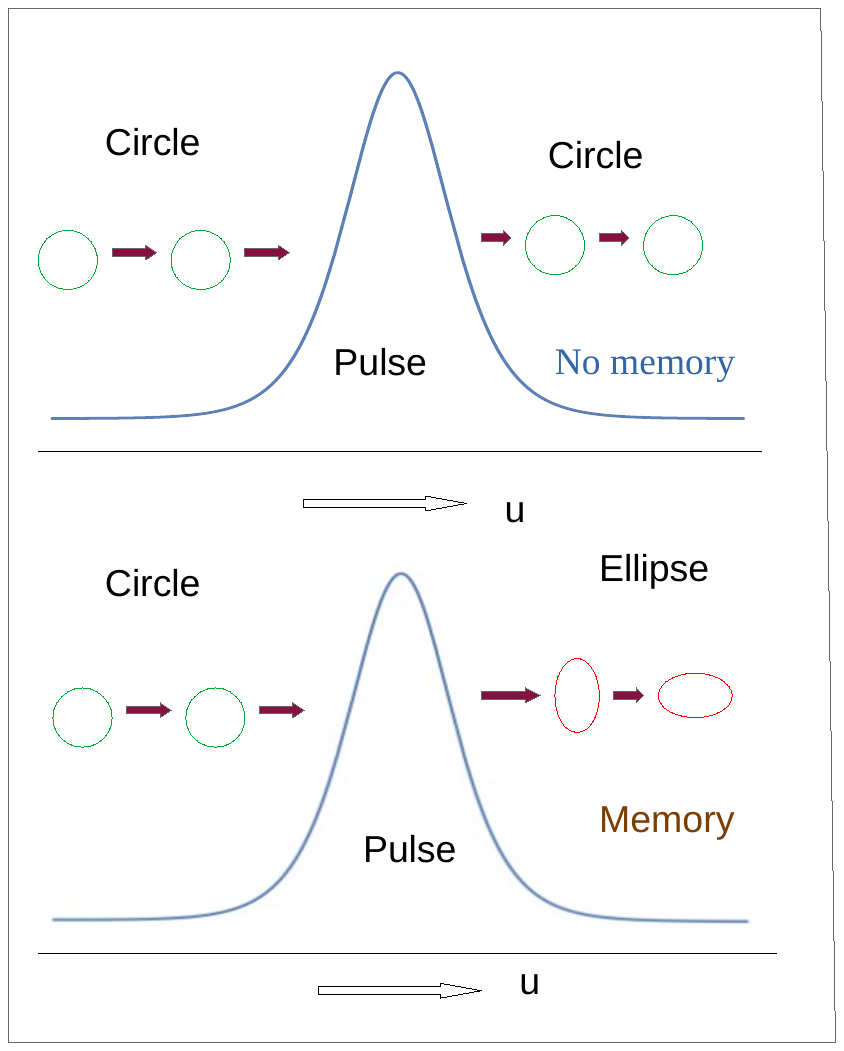}
    \caption{Qualitative idea of memory for a ring of particles experiencing a GW pulse.}
\end{figure}

\noindent If we smoothen out the so-called `ring of particles' what we get is a closed string.
Right now, we are yet to state whether we are dealing with fundamental strings or cosmic strings.
In our way of looking at the problem, it could be anything -- in fact, even a macroscopic, elastic, circular string. The simplest description of such a string, in a relativistic setting, uses the Nambu--Goto area action, the ensuing equations of motion, their solutions and corresponding physical consequences \cite{zweibach}. 
The string configuration $x^i(\tau, \sigma)$, where $\tau$ and $\sigma$ are the worldsheet surface coordinates, is an extremal (minimal) surface in a background spacetime with a metric $g_{ij}(x^k)$. It is given mathematically 
through solutions of the Nambu--Goto equations of motion and constraints. For closed strings, we have the condition--
$x^i(\tau, \sigma) = x^i (\tau, \sigma + 2\pi)$. If such a closed string encounters a gravitational
wave pulse what happens to its circular shape? Do we see a permanent change at later times,
i.e. much after the pulse has disappeared?  We intend to answer such a question in our 
work here.

\noindent How do we model the gravitational wave pulse ? It is possible to use the 
perturbative approach and use a form of $h_{ij}$ which is pulse-like. This may be observationally more relevant. However, here we would like to restrict ourselves to the purely theoretical domain and consider a pp-wave geometry (an exact gravitational wave solution of the vacuum Einstein equations) \cite{ppwave},\cite{blau}. In such a geometry we have a free function which we may choose appropriately, in order to represent a
specific pulse shape.  Hence, we would be studying the evolution of closed strings in a pp-wave spacetime through which we will
try to arrive at what is conventionally known as the `memory' effect. 

\noindent Studies on strings in pp-wave spacetimes are not new.
Past research in this subject area  can be found in 
several older articles \cite{amati, horowitz, devega-sanchez, devega}. Very recently, Li\v{s}ka and von Unge \cite{liska} 
have revisited the topic in an interesting paper which originally motivated us to
pursue the work reported here. On the other hand, string memory 
has been discussed in the recent past in \cite{afshar} and stringy corrections to electromagnetic memory effects in \cite{aldi}.

\noindent In general, gravitational wave memory has been a topic of active research interest over the last few years. The prospect of detection of memory
in future detectors has greatly influenced such studies \cite{memoryexpt0}, \cite{memoryexpt1} (see \cite{memoryexpt3}, \cite{memoryexpt4} and \cite{memoryexpt2} for recent updates). 
The origins of the
idea of GW memory in linearised gravity are in the papers by Zel'dovich, Polnarev \cite{zelpol} and
Braginsky, Grishchuk \cite{bragri} (the term `memory' in the context of gravitational waves was introduced in this paper). Subsequently, Christodoulou\cite{christo}, using full
GR, showed how gravitational waves travelling to null infinity can
give rise to a
contribution to memory. Later, it was {\color[rgb]{0,0,1} claimed} in \cite{biegarf},
using gauge invariant observables, that for linearized gravity with
a null matter stress-energy tensor, one recovers an effect analogous to 
nonlinear memory, though this does not show that nonlinear memory arises in the linearized theory. As emphasized in \cite{biegarf} one still needs to go to quadratic order in general relativity to see nonlinear memory.
Related work on `ordinary' and `null' memory (as defined in \cite{biegarf}) are available in \cite{tolwal,tolwal1, madwin1,madwin2}.
On the other hand, an interesting theoretical link 
involving the memory effects, the Bondi-Metzner-Sachs (BMS) symmetries and soft theorems has been shown to exist and explored in \cite{stro1, stro2}.
Memory in pp-wave and related spacetimes have been extensively studied
in a large number of papers \cite{zhang1,zhang2,zhang3,zhang4,zhang5,ic1, cvet, ic2, ic3, ic4, jpuzan} where diverse aspects of displacement and velocity memory have been
thoroughly analysed. A novel `displacement within velocity memory' which happens for a specific choice of the wave parameters, has recently been reported in \cite{horzhang}. 
A more general approach which aims at defining 
various `persistent gravitational wave observables'  has been pursued by
Flanagan, Nichols, Grant and Harte in a series of papers \cite{eanna1, eanna2, eanna3, eanna4}. An idea named as ${\cal B}$-memory (related to
the ${\cal B}$ tensor which encodes the behaviour of the gradient of the geodesic velocity field) appeared in the
work in \cite{olough} and was followed up in \cite{ic1}. Other work on diverse aspects of memory may be found in \cite{diva, shore, mota}.
The review article \cite{favata1} and the lecture 
by Favata
\cite{favata2} chronicles the history till 2010.  
A very recent review on the memory effect, asymptotic symmetries and its connections with numerical relativity
is available in \cite{cqgfocus1}. \\
\noindent Our article is organised as follows. In the next section (II), we
write down the string equations of motion and constraints in a pp-wave spacetime,
using the conformal gauge. Thereafter, in III, assuming a square pulse gravitational wave,
we obtain solutions at late times, assuming a circular closed string configuration,
before the arrival of the pulse. In this way, via the permanent change of
shape of the closed string we are able to arrive at a memory effect.
We also do a similar analysis for a continuous pulse shape (the sech-squared
pulse). In IV, we discuss some features such as the behaviour of the
induced metric, the worldsheet Ricci scalar and the worldsheet geometry as an
embedding.
Finally in Section V we present our conclusions.

\section{Strings in pp-wave spacetimes}

\noindent Let us begin by first recalling the string equations of
motion and constraints which follow from the variation (w.r.t. $x^i$) of the
Nambu--Goto area action ($S_{NG} = -T_0 \int d\tau d\sigma \sqrt{-\gamma}$,
$\gamma$ being the determinant of the induced metric $\gamma_{ab} =g_{ij} \partial_a x^i \partial_b x^j$ with $a, b...$ being the worldsheet indices $\tau$, $\sigma$ and $T_0$, the string tension). The equations of motion are:
\begin{eqnarray}
\ddot{x}^i-{x^i}''+ \Gamma_{j k}^i \left (\dot{x}^j \dot{x}^k-{x^j}' {x^k}' \right )=0
\end{eqnarray}
and the constraints (conformal gauge) are:
\begin{eqnarray}
g_{ij}\dot{x}^i {x^j}'=0 \\ 
g_{ij}\dot{x}^i \dot{x}^j+g_{ij} {x^i}' {x^j}'=0
\end{eqnarray}
where the dot and prime denote differentiation w.r.t. $\tau$ and $\sigma$, respectively.
The two constraints ensure that the induced metric $\gamma_{ab}$ (as defined above) is
diagonal. We will need to solve the Eqn. (1) written in a background geometry with
metric $g_{ij}$, subject to the constraints in (2) and (3). 

\noindent The plane parallel ($pp$) wave geometry which is a vacuum solution of the
Einstein equations is given by the line element in Brinkmann coordinates as:
\begin{eqnarray}
ds^2=-2du dv+F(u,x,y)du^2+dx^2+dy^2
\end{eqnarray}
where, $u$ and $v$ are null coordinates ($u=\frac{1}{\sqrt{2}} \left(t+z\right )$, $v =\frac{1}{\sqrt{2}} \left (t-z\right )$), $x$ and $y$ are the transverse spatial coordinates and the function $F(u,x,y)$ encodes specific properties of the spacetime. Assuming the above line element as
a solution of the vacuum Einstein equations one ends up requiring $\partial_\alpha \partial^{\alpha} F = 0$ ($\alpha$ being the spatial index representing $x$, $y$).
It is easily seen that 
\begin{eqnarray}
F(u,x,y) = W(u) (x^2-y^2 )
\end{eqnarray}
is a solution for $\partial_\alpha \partial^{\alpha} F = 0$, irrespective of the
form of $W(u)$. It is this free function $W(u)$ which we will choose to specify the
gravitational wave pulse.

\noindent The string equations of motion in the pp-wave spacetime \cite{liska}, for
coordinates $u$, $x$, $y$ are,
\begin{eqnarray}
\left ( \partial_\tau^2-\partial_\sigma^2 \right )u =0  \\
\left (\partial_\tau^2-\partial_\sigma^2 \right ) x = \frac{\partial_x F}{2}\left [(\partial_\tau u)^2-(\partial_\sigma u)^2 \right ] \\
\left ( \partial_\tau^2-\partial_\sigma^2 \right ) y=\frac{\partial_y F}{2}\left [(\partial_\tau u)^2-(\partial_\sigma u)^2 \right ] 
\end{eqnarray}
and the equation for $v$ is
\begin{eqnarray}
\left (\partial_\tau^2-\partial_\sigma^2 \right ) v=\frac{1}{2}\left (\partial_u F\right ) \left [(\partial_\tau u)^2-(\partial_\sigma u)^2 \right ]+ \left (\partial_x F\right ) \left [\partial_\tau u \,\partial_\tau x-\partial_\sigma u\,\partial_\sigma x \right ] \nonumber \\ + \left (\partial_y F\right ) \left [\partial_\tau u\,\partial_\tau y-\partial_\sigma u\,\partial_\sigma y \right ]
\end{eqnarray}
Note that one can solve for the equations for $u$, $x$, $y$ independently and then use the
solutions in the R. H. S. of (9) to write the equation for $v$. The $v$ equation
can then be solved separately.

\noindent A simple solution of the $u$ equation is ($p$, a constant):
\begin{eqnarray}
    u = p \tau
\end{eqnarray}
This identifies the worldsheet time coordinate with the spacetime null coordinate $u$. With this choice, the two gauge conditions in Eqns (2) and (3) reduce to
\begin{eqnarray}
    p \, \partial_\sigma v = \partial_\tau x \, \partial_\sigma x +\partial_\tau y \,
    \partial_\sigma y \\
    2 p \, \partial_\tau v = p^2 F   + \left (\partial_\tau x\right )^2 + \left (\partial_\sigma x\right )^2 + \left  ( \partial_\tau y \right )^2 + \left (\partial_\sigma y\right )^2 
\end{eqnarray}
It can easily be checked that the Eqns. (9), (10), (11) and (12) are mutually consistent and the following equations for $x$ and $y$ hold (using (10)
in (7), (8)):
\begin{eqnarray}
\left (\partial_\tau^2-\partial_\sigma^2 \right ) x = \frac{p^2}{2} \partial_x F\\
\left (\partial_\tau^2-\partial_\sigma^2 \right ) y = \frac{p^2}{2} \partial_y F  
\end{eqnarray}
Therefore, given $F$ one can find $x$ and $y$, which can then be used in
(11) and (12) to find $v$. We will now choose the $W$ in $F$ and try to solve for $x$, $y$ and $v$. Given $F$ in the form in (5) and the assumption
\begin{eqnarray}
    x (\tau, \sigma) = (\cos  \, k_1 \sigma) \, x(\tau) \\
    y(\tau, \sigma ) =  (\sin \, k_1\sigma) \, y(\tau)
\end{eqnarray}
we end up with the following equations for $x(\tau)$ and $y(\tau)$.
\begin{eqnarray}
\ddot x + \left (k_1^2 - p^2 W(\tau) \right ) x =0 \\
\ddot y + \left (k_1^2 + p^2 W(\tau) \right ) y=0
\end{eqnarray}
where the dot denotes differentiation w.r.t. $\tau$ (or, equivalently, $u$ -- except a constant factor `$p$').
In the remote past when $W(\tau)$ is zero we assume
\begin{eqnarray}
x (\tau, \sigma) =  R \cos k_1 \tau \, cos k_1 \sigma \\
y (\tau, \sigma) =  R \cos k_1 \tau \, sin k_1 \sigma 
\end{eqnarray}
which yields a circle with radius $R \cos k_1 \tau$, in the past. We will now assume different types of pulse profiles and solve the Eqns. (17) and (18).
This will lead us to a `permanent change' after the pulse has gone away--a
feature known as the memory effect.

\section{Closed string evolution and memory}

\noindent As mentioned above, we will solve Eqns. (17) and (18) with
specific chosen inputs for $W(\tau)$, representing a pulse. Once the 
solutions for $x(\tau)$ and $y(\tau)$ subject to the initial conditions in
(19), (20) are known, we will use them in (15), (16) to write down the profile functions $x(\tau, \sigma)$
and $y(\tau, \sigma)$. Using the $x(\tau,\sigma)$ and $y(\tau,\sigma)$ in (11), (12)
and integrating once we can obtain the profile $v(\tau,\sigma)$. Knowing
all of $u$, $v$, $x$, $y$ as functions of $\tau$ and $\sigma$ gives us the
embedding of the string world sheet in the background $pp-$wave spacetime.
Let us now proceed with the above scheme.
\subsection{Square pulse}
\begin{figure}[h]
\includegraphics[width=0.6\textwidth]{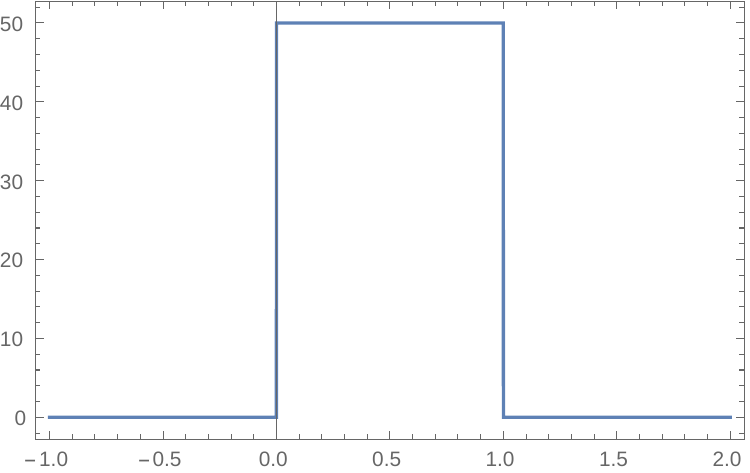}
\caption{The square pulse $W$ as function of $\tau$.}
\end{figure}
\noindent To solve the equations for $x(\tau)$ and $y(\tau)$ we have to assume a functional form for $W(\tau)$. Here we choose it to be the following function (see Figure 2):
\begin{eqnarray}
W(\tau) = 0, \hspace{0.4in} -\infty \le \tau \leq 0
\\ \nonumber
= W_0, \hspace{0.5in} 0\leq \tau \leq T \\ \nonumber
= 0 \hspace{0.9in} \tau \ge T
\end{eqnarray}

\noindent This is the well-known square pulse.
In order to solve the equations for $x(\tau)$ and $y(\tau)$ with the square pulse as
$W(\tau)$ we break up the domain of $\tau$ into three regions ($\tau\leq 0$,
$0\leq \tau\leq T$ and $\tau\geq T$), solve the equations in
the three regions and thereafter match the functions and their derivatives at the two
boundaries $\tau=0$ and $\tau=T$. In the region I ($-\infty \le \tau \leq 0$)
we will assume $x(\tau) = R \cos \, k_1\tau$ and $y(\tau)= R \cos  \, k_1 \tau$. Consequently, it turns out that $x(\tau)= R \cos k_2 \tau$ and
$y(\tau) = R \cos k_3 \tau$ in region II (i.e. between $0\leq \tau\leq T$).

\noindent Finally, after some straightforward calculations, one
arrives at the expressions for $x(\tau)$ and $y(\tau)$ in the $\tau \geq T$ region.
We have, for $x(\tau,\sigma)$ and $y(\tau,\sigma)$:
\begin{eqnarray}
x(\tau,\sigma)= R \left (cos(k_2T)cos[k_1(\tau-T)]-\frac{k_2}{k_1}sin(k_2T)sin[k_1(\tau-T)]\right )cos(k_1\sigma)\\
y(\tau,\sigma)= R \left (cos(k_3T)cos[k_1(\tau-T)]-\frac{k_3}{k_1}sin(k_3T)sin[k_1(\tau-T)]\right )sin(k_1\sigma)
\end{eqnarray}
where,
\begin{eqnarray}
k_2 = \sqrt{k_1^2 - W_0 p^2} \hspace{0.2in}; \hspace{0.2in}
k_3 = \sqrt{k_1^2 + W_0 p^2}
\end{eqnarray}
It is easily evident from the expressions in (22) and (23) that the
profile (in the $xy$ plane) at any $\tau \geq T$ is not a circle, even though it was a circle of
radius `$R \cos \, k_1\tau$' before $\tau=0$. For example, if we look at the profile
at any value $\tau= T+ \frac{2n \pi}{k_1}$ ($n=0,1,2....$) we always get
\begin{eqnarray}
x = R \cos (k_2 T) \, \cos k_1 \sigma \hspace{0.2in};\hspace{0.2in}
y = R \cos (k_3 T) \, \sin k_1 \sigma 
\end{eqnarray}
Since $k_2$ and $k_3$ are never equal for $W_0\neq 0$, the profile of the
closed string  in the $xy$ plane, is an ellipse. One may track the shape of the closed string
for different $\tau\geq T$ and obtain the behaviour with the progress of
$\tau$. It is easy to see that the ellipse changes its shape with the evolution
of $\tau$. It does degenerate to a line along the $x$ axis when the $y$ coordinate
becomes zero and vice versa. However, as is seen from the expression, it is never
possible for both $x$ and $y$ to be zero  or equal simultaneously. Therefore, we
never see a circular string again. This is the `permanent change' which is akin to
a {\em memory effect caused by a gravitational wave pulse}.
Figure 3 shows the string profile in the $xy$ plane, at different increasing $\tau \geq T$ values,
in order to illustrate the permanent deformation of the closed circular string.
Notice that the shape of the pulse plays a crucial role in creating this deformation, through the product $W_0 T$ (height times the width).
\begin{figure}[h]
         \includegraphics[width=0.6\textwidth]{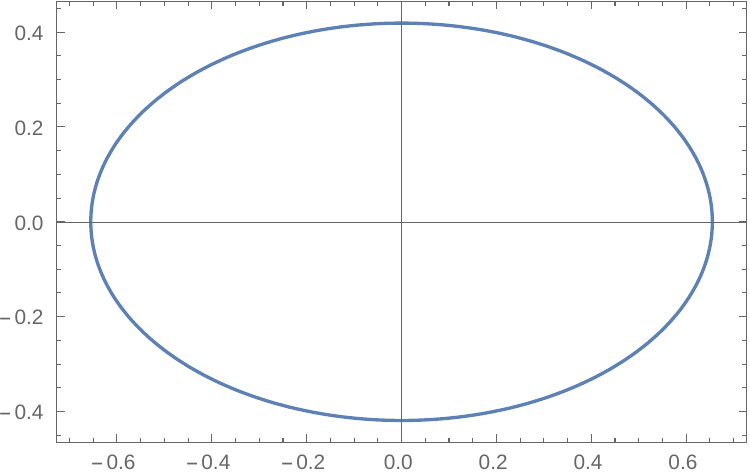}
         \includegraphics[width=0.3\textwidth]{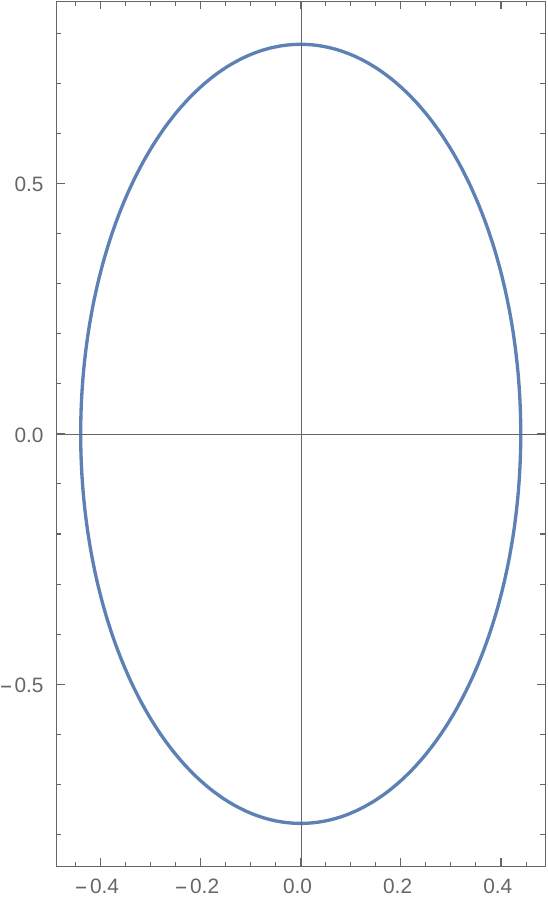}
    \caption{The string profile in the $xy$ plane for a square pulse, at two
    different values of $u$ or $\tau$, after the pulse has left. The $v$ coordinate
    is suppressed here. We have chosen 
    $T=1$ unit, $k_1T=1$, $k_2 T = 0.5$, $k_3 T = \frac{\sqrt{7}}{2}$, $R=1$ unit. The $\tau$ values
    are $\tau=1.5$ unit  (left) and a later value, $\tau=1.8$ unit (right)}
\end{figure}

\noindent It is possible to rewrite the above discussion in a general matrix notation. Let us consider column vectors ${\bf X} =(x\,\,\, y)$, ${\bf Q} =(cos\,k_1\sigma\,\, \, sin\,k_1\sigma)$ and write the above $x$, $y$ profiles as a matrix equation,
\begin{eqnarray}
{\bf X} = R \, {\bf A^{-1}Q}
\end{eqnarray}
where ${\bf A^{-1}}$ is the coefficient matrix. For the example discussed just above, this matrix has diagonal elements which can be read off from (22) and (23). In general, we obtain the relation,
\begin{eqnarray}
{\bf X^{T} \left ( A^TA \right ) X}=R^2
\end{eqnarray}
The string will be a circle if 
\begin{eqnarray}
{\bf A^T A}=h(\tau){\bf I}
\end{eqnarray}
For $\tau < 0$, we get $h(\tau)=(cos \, k_1 \tau)^{-2}$. But, for $\tau\geq T$,
${\bf A^T A}$ is not proportional to the identity matrix, which indicates the
presence of a memory effect.

\noindent In order to complete this exercise, we must know the $v$ coordinate too.
This is a somewhat involved calculation and we may set $p= \frac{k_1 R}{\sqrt{2}}$ ( as long as we choose to assume $u=v$ before the arrival of the pulse), to obtain an expression for $v$ which will satisfy (11) and (12)
with the inputs for $x$ and $y$ from (22) and (23). We end up with the following expressions in the three regions.

\noindent {\bf Region $\tau\leq 0:$}
\begin{eqnarray}
u=\frac{k_1 R}{\sqrt{2}} \tau \hspace{0.2in}; \hspace{0.2in} v= \frac{k_1 R}{\sqrt{2}}\tau
\\
x = R\cos k_1 \tau \cos k_1 \sigma \hspace{0.2in} ; \hspace{0.2in} y = R \cos k_1 \tau \sin k_1 \sigma
\end{eqnarray}

\noindent {\bf Region $0\leq \tau\leq T$:}
\begin{eqnarray}
u=\frac{k_1 R}{\sqrt{2}} \tau \hspace{0.2in}; \hspace{0.2in}
x = R\cos k_2\tau \cos k_1 \sigma \hspace{0.2in} ; \hspace{0.2in} y = R \cos k_3 \tau \sin k_1 \sigma \\
v (\tau, \sigma) = -\frac{R}{2\sqrt{2}} \frac{\cos 2 k_1 \sigma}{k_1^2} G_0(\tau)
+ H_0(\tau) 
\end{eqnarray}
where,
\begin{eqnarray}
G_0(\tau) =  k_1 k_2 \cos k_2 \tau \sin k_2 \tau - k_1 k_3 \cos k_3 \tau \sin k_3 \tau \nonumber
\end{eqnarray}
and
\begin{eqnarray}
H_0(\tau) =
\frac{k_1 R}{\sqrt{2}} \tau + \frac{k_1^2-k_2^2} {2\sqrt{2} k_1} R \left (\frac{\sin (2k_2 \tau)}{2k_2} - \frac{\sin (2k_3 \tau)}{2k_3}\right )
\end{eqnarray}

\noindent {\bf Region $\tau\geq T$:}
\begin{eqnarray}
u=\frac{k_1 R}{\sqrt{2}} \tau \hspace{0.2in}; \hspace{0.2in}
v (\tau, \sigma) = -\frac{R}{2\sqrt{2}} \frac{\cos 2 k_1 \sigma}{k_1^2} G(\tau)
+ H(\tau) 
\end{eqnarray}
where,
\begin{eqnarray}
G(\tau) = \frac{1}{2} \left (k_1^2 \cos^2 k_2 T - k_2^2 \sin^2 k_2 T \right )
\sin \, 2 k_1 (\tau-T) + k_1 k_2 \cos k_2 T \sin k_2 T \cos 2 k_1 (\tau-T) \nonumber \\
- \frac{1}{2} \left (k_1^2 \cos^2 k_3 T - k_3^2 \sin^2 k_3 T \right )
\sin \, 2 k_1 (\tau-T) - k_1 k_3 \cos k_3 T \sin k_3 T \cos 2 k_1 (\tau-T) \nonumber
\\
\end{eqnarray}
and
\begin{eqnarray}
H(\tau) = \frac{R}{2\sqrt{2} k_1} \left ( k_1^2 \cos^2 k_2 T + k_2^2 \sin^2 k_2 T
+k_1^2 \cos^2 k_3 T + k_3^2 \sin^2 k_3 T \right ) \left (\tau - T\right ) + \nonumber \\
\frac{k_1 R}{\sqrt{2}} T + \frac{k_1^2-k_2^2} {2\sqrt{2} k_1} R \left (\frac{\sin (2k_2 T)}{2k_2} - \frac{\sin (2k_3 T)}{2k_3}\right )
\end{eqnarray}
with $x$ and $y$ for $\tau\geq T$ as given in (22), (23).

\noindent It can be verified that the above forms of $x$, $y$, $u$ and $v$ match at
the boundaries $\tau=0$ and $\tau=T$. The $v$ coordinate exhibits
a derivative discontinuity at the boundaries because of the piece-wise
continuous nature of the square pulse.

\noindent The above analysis shows that before the pulse arrives, the centre of mass
of the string (given via $z= \frac{1}{\sqrt{2}} (u-v)$) is at $z=0$. However,
after the pulse arrives and leaves, the centre of mass develops a translational
motion in $\tau$ as well as an oscillation. One can go over to a Lorentz frame
(Lorentz transformation in the $t,z$ coordinates) and make sure that the centre of mass
has only oscillatory motion after the pulse leaves.  With such a Lorentz transformation
the new $u$, $v$ coordinates (named as $u'$, $v'$) become:
\begin{eqnarray}
u' = \sqrt{a} \frac{k_1 R}{\sqrt{2}} \tau \\
v' = \sqrt{a} \frac{k_1 R}{\sqrt{2}} \tau + \frac{k_1 R\, b}{\sqrt{2 a}} - \frac{R}{2\sqrt{2a}}\frac{\cos 2k_1\sigma}{k_1^2} G(\tau) 
\end{eqnarray}
where $a$ and $b$ are, as defined in $H(\tau) = \frac{k_1 R}{\sqrt{2}} \left (a\tau+ b\right  )$ with 
\begin{eqnarray}
a = \frac{1}{2k_1^2} \left ( k_1^2 \cos^2 k_2 T + k_2^2 \sin^2 k_2 T
+k_1^2 \cos^2 k_3 T + k_3^2 \sin^2 k_3 T \right )
\\
b = - \frac{1}{2 k_1^2} \left ( k_1^2 \cos^2 k_2 T + k_2^2 \sin^2 k_2 T
+k_1^2 \cos^2 k_3 T + k_3^2 \sin^2 k_3 T \right ) T  + \nonumber \\
T + \frac{k_1^2-k_2^2} {2 k_1^2} \left (\frac{\sin (2k_2 T)}{2k_2} - \frac{\sin (2k_3 T)}{2k_3}\right )
\end{eqnarray}
and the magnitude of the boost velocity between the two frames is given as $v_{boost} = 
\frac{1-a}{1+a}$. For $a>1$, $v_{boost} \rightarrow -v_{boost}$ and
one needs to replace $a$ with $\frac{1}{a}$ in the above formulae.
Further, from $u'$ and $v'$ one can obtain $t'$ and $z'$.

\noindent The emerging picture of the closed string before and after it experiences the pulse
is as follows:

\noindent $\bullet$ Before the pulse arrives the closed string shape is circular and its centre of mass is at $z=0$. The radius of the circle changes with $\tau$ (a pulsating string) and
can even become zero at specific $\tau$ values. The $v$ coordinate 
is chosen to have the same value
as the $u$ coordinate, for any $\tau$. In fact $u=v$ as in Eqn. (29).

\noindent $\bullet$ After the pulse departs, the closed string profile
in the $xy$ plane (at a fixed $\tau$ value) is elliptical in shape
and never becomes circular. If we include the behaviour of the $v$ coordinate, we can see from (32) or (34) that for fixed $\tau$, $v$ has different values at different $\sigma$. In fact, at fixed $\tau$, one can relate the spacetime coordinate $v$ with
the worldsheet coordinate $\sigma$, for the given embedding. The shape of the string can degenerate to a line (along $x$ or $y$) at
specific $\tau$ values. The centre of mass ($z'= \frac{1}{\sqrt{2}} (u'-v')$ value) is now purely oscillatory
with different amplitudes at different $\tau$ (construct $z'$ from (37), (38)) -- it does not have
any linear in $\tau$ motion as long as we are in a properly chosen frame. Note that
prior to the arrival of the pulse, the centre of mass was at $z=0$ (see Eqn. (29)). The string profiles in $xyz'$ space at chosen values of
$\tau$ are shown in Figure 4. One can see that before the pulse arrives, the string is confined to the $z=0$ plane, though later, it
spreads out.  
\begin{figure}[h]
         \includegraphics[width=0.6\textwidth]{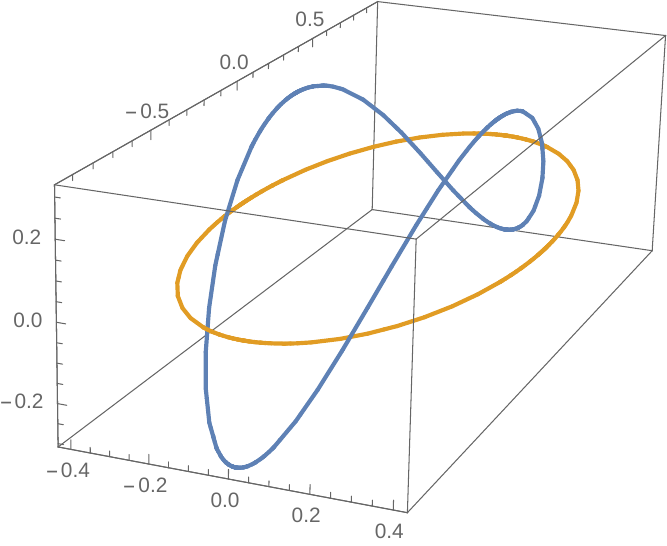}
        \caption{The string profile in $xyz'$ space for a square pulse, at two
    different values of $u$ or $\tau$, before the arrival (yellow curve) and after the departure (blue curve) of the pulse. We have chosen 
    $T=1$ unit, $k_1T=1$, $k_2 T = 0.5$, $k_3 T = \frac{\sqrt{7}}{2}$, $R=1$ unit. Eqns (22), (23) and (37)-(40) are used.  The $\tau$ values
    are $\tau=-2$ unit  (yellow) and a later value, $\tau=2$ unit (blue). The vertical axis is the $z'$ axis.}
\end{figure}

\noindent There are changes in the geometry of the worldsheet caused by the pulse.
We will discuss these aspects briefly in the penultimate section of this article.

\subsection{Sech-squared pulse}

\noindent It is easy to redo the above analysis for a delta function pulse
($W(\tau)= W_0 \delta (\tau)$). The results are similar to the square pulse
and the mathematical details have been worked out in \cite{liska}, though the
authors in \cite{liska} unfortunately missed mentioning the `memory' aspect! 
The square and the delta function pulses are largely idealistic. We therefore
move on now to a pulse shape which is smooth--the so--called sech-squared pulse.
The $W(\tau)$ here is given as (see Figure 5):
\begin{eqnarray}
W(\tau) = \frac{\alpha^2}{4p^2} sech^2 \alpha \tau 
\end{eqnarray}
\begin{figure}[h]
         \includegraphics[width=0.7\textwidth]{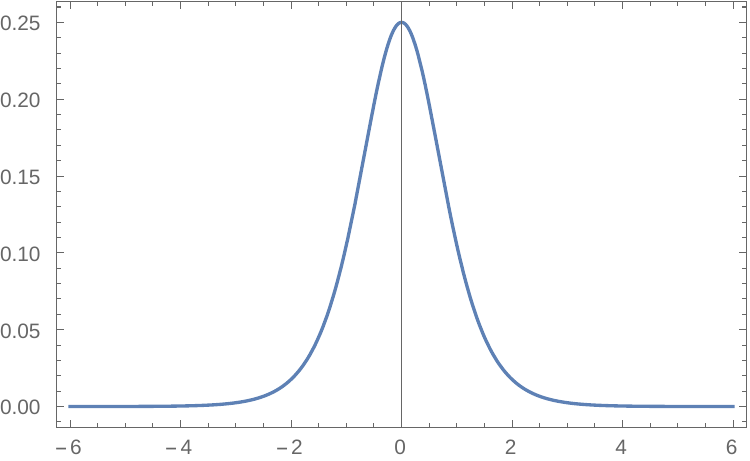}
    \caption{The Sech-squared pulse $W$ as function of $\tau$ (or $u$). Here $\alpha=1$.}
\end{figure}
The two differential equations which we need to solve to obtain $x(\tau)$ 
and $y(\tau)$ are,
\begin{eqnarray}
\ddot x + \left (k_1^2 -  \frac{\alpha^2}{4} sech^2 \alpha \tau  \right ) x =0 \\
\ddot y + \left (k_1^2 +  \frac{\alpha^2}{4} sech^2 \alpha \tau  \right ) y=0
\end{eqnarray}
Both these equations can be solved exactly and the asymptotic forms of the
solutions are known. As we did earlier, we will impose the circular closed string shape at negative infinity (i.e. $\tau \rightarrow - \infty$). Using the known
asymptotic forms we will then find out the shape of the closed string as $\tau \rightarrow \infty$. A change in the shape, like before, will be associated with
a permanent deformation and a memory effect caused by the pulse.

\noindent The Eqns. (42) and (43) fall in the general class of Poschl--Teller potential
problems usually studied in quantum mechanics \cite{cevik}. The two linearly independent solutions for
equations of the type 
\begin{eqnarray}
\ddot U(\tau) + \left ( k^2 + \frac{\alpha^2 \lambda(\lambda -1)}{\cosh^2 \alpha \tau}\right ) U (\tau) =0
\end{eqnarray}
are given as
\begin{eqnarray}
U_1(\tau) =  \left (1+\tanh \alpha \tau \right )^{\frac{ik}{2\alpha}}  \left (1-\tanh \alpha \tau \right )^{-\frac{ik}{2\alpha}} {}^2 F_{1} \left ( \lambda; 1-\lambda; \frac{ik}{\alpha} +1; \frac{1+\tanh \alpha \tau}{2} \right ) \nonumber \\
U_2 (\tau) = 2^{\frac{ik}{\alpha}} \left (1-\tanh^2 \alpha \tau \right )^{-\frac{ik}{2\alpha}} {}^2 F_{1} \left ( \lambda -\frac{ik}{\alpha}; 1-\lambda - \frac{ik}{\alpha}; 1-\frac{ik}{\alpha}; \frac{1+\tanh \alpha \tau}{2} \right )
\end{eqnarray}
where ${}^2 F_1$ denotes the hypergeometric function. 
The general solution therefore is,
\begin{eqnarray}
U(\tau) =  A U_1 (\tau) + B U_2 (\tau)
\end{eqnarray}
where $A$ and $B$ are constants to be determined by the initial conditions. 
Eqn. (44) will match with (42) when $\lambda = \frac{1}{2}$. Similarly, when   $\lambda = \frac{1+\sqrt{2}}{2}$ we will get Eqn. (43).

\noindent We now need to write down the asymptotic form of the general solution in
(45) for $\tau \rightarrow -\infty$ and $\tau \rightarrow +\infty$.

\noindent Notice that for $\tau \rightarrow -\infty$, the argument of the
hypergeometric function goes to zero and the value of ${}^2F_1 (a;b;c;0)$ is equal to one. Hence, expanding the hyperbolic tangents as exponentials and taking the
limit $\tau \rightarrow -\infty$ we have
\begin{eqnarray}
U (\tau \rightarrow -\infty) = A e^{i k \tau} + B e^{-ik \tau}
\end{eqnarray}
The limit of $\tau \rightarrow +\infty$ is a little tricky to obtain.
One will have to use a linear transformation formula  \cite{abramowitz}, \footnote{Use is 
made of the formula 15.3.6, Pg. 559 in \cite{abramowitz}.} and write the hypergeometric function
with the argument $z=\frac{1+\tanh \alpha \tau}{2}$ in terms of a sum (with a
$z$ dependent coefficient) of two
hypergeometric functions with arguments $1-z=\frac{1-\tanh \alpha \tau}{2}$.
This helps in getting the asymptotic form easily because, now, as one takes the limit $\tau\rightarrow \infty$
one ends up with the argument as zero in the two hypergeometric functions and their value become equal to one. 

\noindent Thus, as $\tau \rightarrow \infty$ one gets
\begin{eqnarray}
U(\tau \rightarrow \infty) = A' e^{ik \tau} + B' e^{-ik \tau}
\end{eqnarray}
where
\begin{eqnarray}
 A' =  \frac{\Gamma (\frac{ik}{\alpha} +1)  \Gamma (\frac{ik}{\alpha})}{\Gamma (\frac{ik}{\alpha}+1-\lambda) \Gamma (\frac{ik}{\alpha} +\lambda)} \, A + 
 \frac{\Gamma (1-\frac{ik}{\alpha})  \Gamma (\frac{ik}{\alpha})}{\Gamma (1-\lambda) \Gamma (\lambda)} \, B  =  a_1 \, A  + b_1 \, B
\end{eqnarray}
\begin{eqnarray}
    B' = \frac{\Gamma (\frac{ik}{\alpha} +1)  \Gamma (-\frac{ik}{\alpha})}{\Gamma (1-\lambda) \Gamma (\lambda)} \, A + 
 \frac{\Gamma (1-\frac{ik}{\alpha})  \Gamma (-\frac{ik}{\alpha})}{\Gamma (\lambda -\frac{ik}{\alpha}) \Gamma (1-\lambda-\frac{ik}{\alpha})} \, B  =  a_2 \, A  + b_2 \, B
\end{eqnarray}
Armed with these asymptotic forms we will now write down the solutions for
Eqns. (42) and (43).

\noindent Let us first consider Eqn. (42), i.e. the $x$ equation.
We keep in mind that we need to use $\lambda=\frac{1}{2}$ while using the
solutions quoted above.
Here, as $\tau \rightarrow - \infty$ we need $x(\tau) = R \cos k_1 \tau$
(and $x(\tau,\sigma) = R \cos k_1 \tau \cos k_1 \sigma$).
Therefore, one has to choose $A = B = \frac{R}{2}$. With this asymptotic condition
at $\tau \rightarrow -\infty$, one can easily write down the 
final solution for $\tau \rightarrow \infty$. This is given as (taking the real
part of the full complex solution, since $a_1, b_1, a_2, b_2$ can be complex):
\begin{eqnarray}
    x(\tau, \sigma) = \frac{R}{2} Re \left [ \left (a_1+b_1+a_2+b_2 \right )
    \cos k_1\tau \cos k_1 \sigma + \left (a_2+b_2-a_1-b_1\right ) \sin k_1 \tau \sin k_1\sigma \nonumber \right . \\ + \left . i \left (a_1+b_1-a_2-b_2 \right ) \sin k_1\tau \cos k_1\sigma + i \left (a_1+b_1+a_2+b_2\right )\cos k_1\tau \sin k_1\sigma \right  ] \nonumber \\
    \end{eqnarray}

\noindent For Eqn. (43), we need to use $\lambda = \frac{1+\sqrt{2}}{2}$. 
The $\tau\rightarrow -\infty$ asymptotic form of  $y (\tau,\sigma)$ is given as $R\cos k_1 \tau \sin k_1\sigma$. To implement this form in the asymptotic past, we will need to use $A=B=\frac{R}{2i}$ in the solutions
quoted above. We will write the final results using $a_1'$, $b_1'$, $a_2'$ and
$b_2'$ since we have a different equation and a different $\lambda$ value.
Thus, we have $y(\tau,\sigma)$ as:
\begin{eqnarray}
    y(\tau, \sigma) = \frac{R}{2} Re \left [ -i \left (a_1'+b_1'+a_2'+b_2' \right )
    \cos k_1\tau \cos k_1 \sigma - i \left (a_2'+b_2'-a_1'-b_1'\right ) \sin k_1 \tau \sin k_1\sigma \nonumber \right . \\ + \left .  \left (a_1'+b_1'-a_2'-b_2' \right ) \sin k_1\tau \cos k_1\sigma +  \left (a_1'+b_1'+a_2'+b_2'\right )\cos k_1\tau \sin k_1\sigma \right ] \nonumber \\
    \end{eqnarray}

\noindent It is evident from the above expressions that the circular string in the
asymptotic past, under the influence of the gravitational wave pulse will change
its shape and become elliptical in shape, in the asymptotic future. 
To see this explicitly, we will now work out the details with explicit values for the
various parameters 
and then plot the profiles. We have chosen to retain values upto four decimal places.

\noindent Let us consider the case where $k_1=1$, $\alpha=1$.
Using the expressions $a_1$, $a_2$, $b_1$, $b_2$ with $\lambda=\frac{1}{2}$ we obtain
\begin{eqnarray}
x(\tau,\sigma) = \frac{R}{2} \left ( 1.9379 \cos \, \tau + 0.6971 \sin \, \tau \right ) \cos \, \sigma 
\end{eqnarray}
For $\lambda = \frac{1+\sqrt{2}}{2}$ and the same $k_1$, $\alpha$
one gets, using $a_1'$, $b_1'$, $a_2'$ and $b_2'$,
\begin{eqnarray}
y(\tau, \sigma) = \frac{R}{2} \left ( 1.9461 \cos \, \tau - 0.5777 \sin \, \tau \right ) \sin \, \sigma 
\end{eqnarray}
\begin{figure}[h]
         \includegraphics[width=0.38\textwidth]{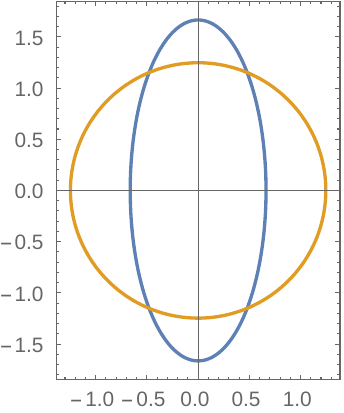}
         \includegraphics[width=0.55\textwidth]{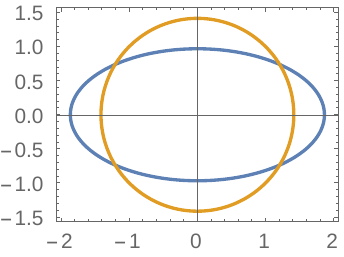}
        \caption{The late time string configurations (blue curves) at $\tau= \frac{103 \pi}{7}$
        (left) and at $\tau = \frac{101 \pi}{4}$ (right). The yellow circles 
        correspond to the earlier configurations at $\tau= - \frac{103 \pi}{7}$ and
        $\tau= -\frac{101 \pi}{4}$. We have assumed $R=2$ units and Eqns.
        (19), (20) and (53), (54). }
        \label{fig:SgrAG}
\end{figure}

\noindent The expressions are, by definition, the profiles at large values of
$\tau$. However, we can get an idea of the elliptical shape
by plotting at a large, finite value of $\tau$. Two representative plots
are shown in Figure 6. The initial profile (at $\tau \rightarrow -\infty$) is just a circle 
( $x = 2 \cos \tau \cos \sigma$, $y= 2\cos \tau \sin \sigma$). 

\noindent A permanent change in the shape of the string appears
to have taken place. The cause of this is, like in the square pulse case, the gravitational wave pulse -- given here as a sech-squared
function of $\tau$  or $u$.

\noindent It is possible to obtain $v(\tau, \sigma)$ in this case
too though we do not attempt to analyse it here. Qualitatively, the result
is not expected to be very different from the square pulse case
discussed earlier.

\section{Worldsheet geometry and memory}
\noindent Let us now focus on the geometry of the worldsheet. We will
consider here the case of the square pulse only. The worldsheet
line element can be found using the embedding functions and the
worldsheet metric functions $\gamma_{ab} = g_{ij} \partial_a x^i \partial_b x^j$.
As expected, the line element is diagonal (conformal gauge choice).
We have, for $\tau\leq 0$, the line element as given by,
\begin{eqnarray}
ds_{\tau\leq 0}^2 =  R^2k_1^2\cos^2 \, k_1 \tau \left (-d\tau^2 +d\sigma^2\right )
\end{eqnarray}
The metric is degenerate (zero determinant) 
at all $\tau = \frac{(2n+1)\pi}{2k_1}$ ($n=0,1,2....)$.

\noindent On the other hand the line element for $\tau\geq T$ is given as
\begin{eqnarray}
ds^2 = \Omega^2 (\tau,\sigma) \left (-d\tau^2 + d\sigma^2\right )
\end{eqnarray}
where
\begin{eqnarray}
    \Omega^2 (\tau,\sigma) = R^2 \sin^2 k_1\sigma \left [ k_1 \cos\, k_2 T \cos \,
    k_1 (\tau-T) - k_2 \sin \, k_2 T \sin k_1 (\tau-T) \right ]^2 + \nonumber \\
    R^2 \cos^2 k_1\sigma \left [ k_1 \cos\, k_3 T \cos \,
    k_1 (\tau-T) - k_3 \sin \, k_3 T \sin k_1 (\tau-T) \right ]^2
\end{eqnarray}
The metric here is never degenerate if $\Omega^2 (\tau, \sigma)$ is not zero for any value of $\tau$, $\sigma$. How do we check this? Since $\Omega^2$ is a sum of two squares,
it can only be zero if the individual terms are both zero. This
seems possible, say at $\tau=\tau_c$, if $k_2 \tan \, k_2 T = k_3 \tan k_3 T = \frac{k_1}{\nu}$, where $\tan k_1 (\tau_c-T)=\nu$. 
Further, we have $k_2 < k_3$ and $k_2^2+k_3^2 = 2k_1^2$. 
It is possible to solve the first equation, i.e.  $k_2 \tan \, k_2 T = k_3 \tan k_3 T$, 
to arrive at a value of $T$ (given $k_1$, $k_2$). This value of $T$ can then be used in the
other equation, i.e. $k_2 \tan \, k_2 T = k_1 \cot k_1 (\tau_c-T)$, to obtain $\tau_c$.
For example, assuming, in appropriate units, $k_1=\frac{5}{\sqrt{2}}$, $k_2=3$ and $k_3=4$, we can easily see that $T= m\pi$ solves the first equation. Using $m=1$, we get 
one value $\tau_c = \pi (1+ \frac{\sqrt{2}}{10})$ (others are there too) from the second equation. 
In fact, $T=m\pi$ will always satisfy the first equation as long as $k_2$, $k_3$ are integers satisfying the previously stated constraints. Hence, at such $\tau$ values, for the specific $T$, one does get $\Omega^2=0$, but, for all other $T$ one can have $\Omega^2\neq 0$ everywhere. Therefore, if the pulse is such that its width
$T$ is different from the set of $T$
values which yield $\Omega^2=0$ at some $\tau=\tau_c$, one does indeed end up with a non-degenerate metric.

\noindent The Ricci scalar ${\cal R}$ of the two dimensional worldsheet also
undergoes a change in character after the passage of the pulse. Recall that
the expression for the Ricci scalar in terms of the conformal factor $\Omega^2$ is
\begin{eqnarray}
{\cal R} = -\frac{2}{\Omega^2} \left (-\frac{\partial^2}{\partial \tau^2} +
\frac{\partial^2}{\partial\sigma^2} \right ) \ln \Omega
\end{eqnarray}
The denominator factor in the above expression can lead to a worldsheet singularity
where $\Omega^2=0$. Thus, singularities always exist prior to the arrival of the
pulse, but  may get removed (depending on the  value of $T$ for the pulse) after its passage. 

\noindent Both the above features-- (a) metric becoming non-degenerate
everywhere and (b) the worldsheet singularites being removed --- are `permanent changes' which could possibly be
caused by the gravitational wave pulse. Thus, apart from the change of shape of
the circular string, we may also see changes in the worldsheet
geometry which seem to add on to the character of `memory'.

\begin{figure}[h]
\includegraphics[width=0.3\textwidth]{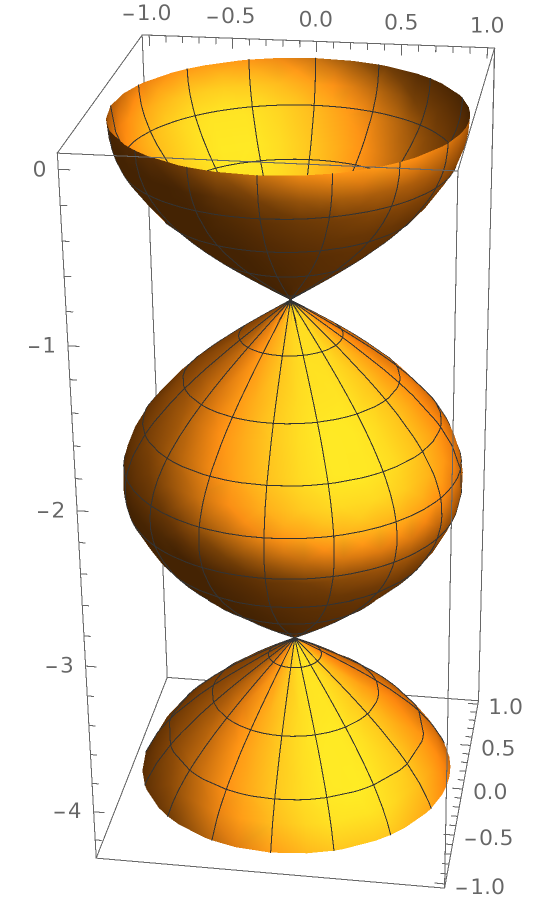}
\caption{The worldsheet geometry before the arrival of the pulse. The zero
radius locations are the worldsheet singularities. We have used Eqns. (29), (30)
with $R=1$. $k_1\tau$ and $k_1\sigma$ are the surface coordinates.}
\end{figure}
\begin{figure}[h]
\includegraphics[width=0.35\textwidth]{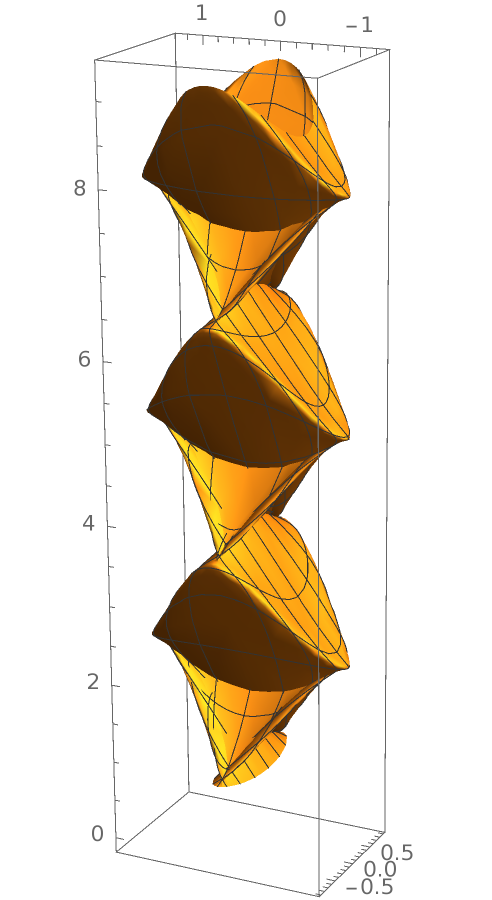}
\includegraphics[width=0.25\textwidth]{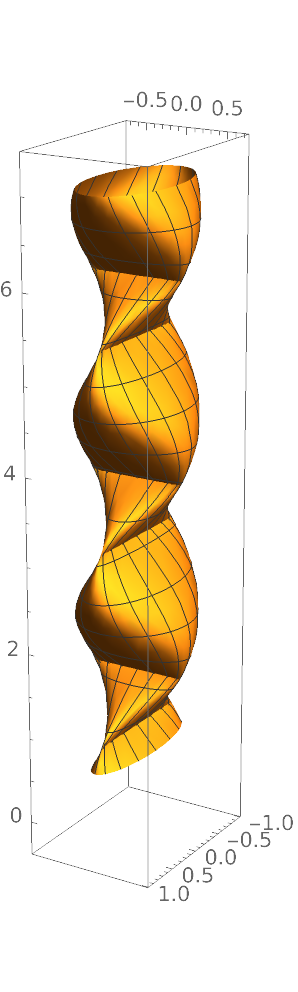}
\caption{The worldsheet geometry as a $xyv$ embedding (left) and
a $xyu$ embedding (right) after the departure of the pulse. There are no locations where
the worldsheet shrinks to a point. Hence, no worldsheet singularities. Eqns. (22), (23) and (34)-(36) are used with values for the parameters $T$, $k_1 T$,
$k_2 T$, $k_3 T$ and $R$ as mentioned for Figure 3. The surface coordinates are $k_1\tau$ and $k_1\sigma$.}
\end{figure}

\noindent One can visualise the two dimensional surface of the worldsheet
by embedding  it in a three dimensional Euclidean background. The string worldsheet however is a two dimensional surface in the four dimensional pp-wave spacetime, 
which, as we noted before, is just flat spacetime before and after the duration of the pulse. Embeddings as hypersurfaces in Euclidean space 
can therefore provide only a partial visualisation. We show the embedded (in a three dimensional background) sections of the  
worldsheet geometry in Figures 7-9. 

\noindent Figure 7
shows the embedding 
($x(\tau,\sigma), y(\tau,\sigma), u(\tau) = \frac{1}{\sqrt{2}} \tau$) before the arrival of the pulse. One can see the presence of
singularities on the worldsheet at locations where it collapses to a point.
As mentioned before, it is a pulsating string. Here, i.e. prior to the arrival of the pulse, the embedding ($x(\tau,\sigma), y(\tau,\sigma), v(\tau) = \frac{1}{\sqrt{2}} \tau$) is the same as the $xyu$ embedding.

\noindent Figure 8 shows the worldsheet after the pulse departs. One can see  here that
the worldsheet is indeed nonsingular -- it never collapses to a point--at best
it can collapse to a line.  The plot (left) is obtained by using the
parametric relations for $x(\tau,\sigma)$, $y(\tau,\sigma)$ and $v(\tau, \sigma)$. Since $u$ is proportional to $\tau$, the spacetime null coordinate $u$ is proportional to the 
worldsheet `time' $\tau$. Before the pulse arrives $u=v$, but later it is not so. The embedding $x(\tau,\sigma)$, $y(\tau,\sigma)$ and $u(\tau)$ is shown on the right, in Figure 8, where the nonsingular nature of the worldsheet is perhaps more clearly visible.
For the embedding $xyv$, at any $\tau$ value, points at different $\sigma$
do not lie on a plane and the string is spread away from any $u=$constant plane. This does not happen for the $xyu$ embedding.
Further, one can construct $xyz'$ and $xyt'$ embeddings which are
shown in Figure 9. One may note the connection between the blue curve
(at one later $\tau$ value)
in Figure 4 and the finite zone (over the range for $\tau$) 
within which the string wraps itself while evolving (as displayed in the $xyz'$ embedding), in Figure 9. 
\begin{figure}[h]
\includegraphics[width=0.35\textwidth]{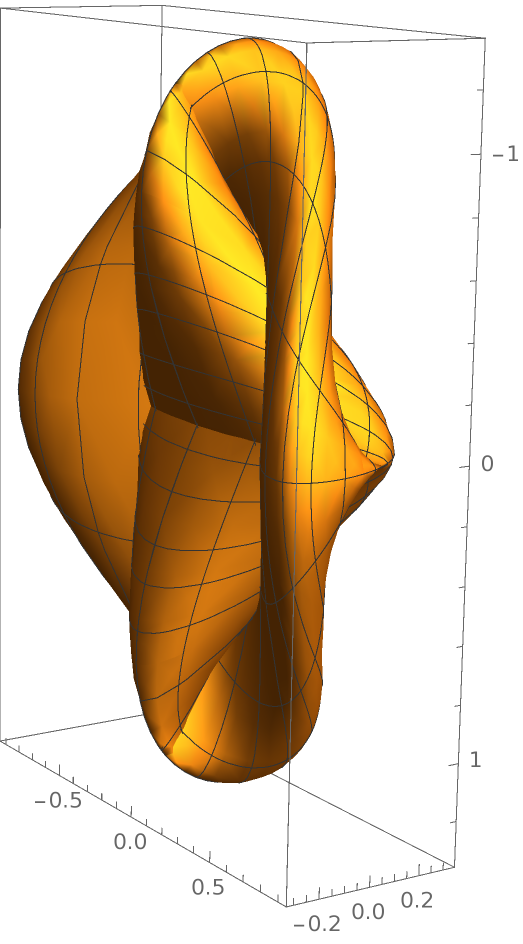}
\includegraphics[width=0.40\textwidth]{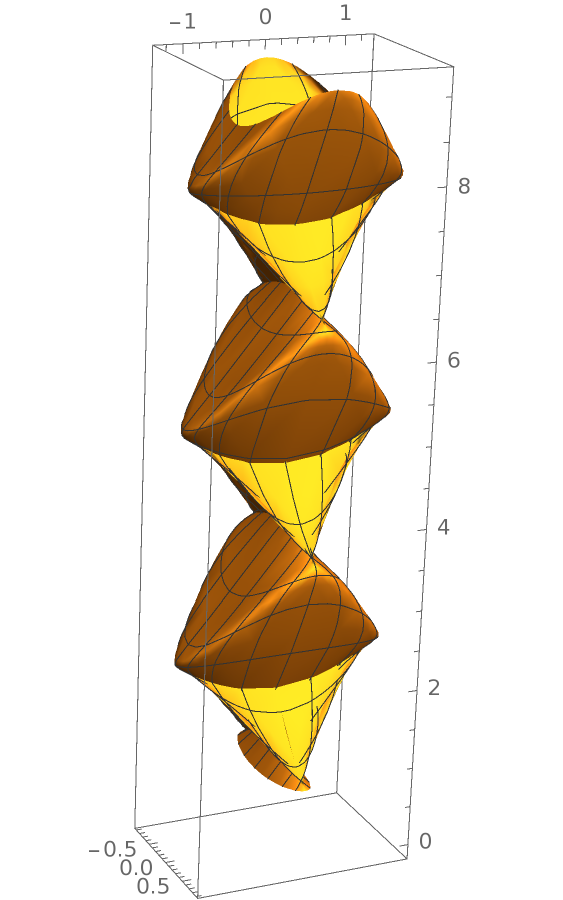}
\caption{The worldsheet geometry as a $xyz'$ embedding (left)
and a $xyt'$ embedding (right), after the arrival of the pulse.  Eqns. (22), (23) and expressions for $t'$ and $z'$(derived from Eqns. (37) - (40)) are used with values for the parameters $T$, $k_1 T$,
$k_2 T$, $k_3 T$ and $R$, as mentioned in Figure 3. The surface coordinates are $k_1\tau$ and $k_1\sigma$.}
\end{figure}

\noindent All the above `visualisations' are of some use in our attempt to understand the nature of the worldsheet in the four dimensional spacetime. As stated earlier, since the string worldsheet is a codimension two surface in a four dimensional background
one needs to look at different embeddings in order to develop a 
clear picture. Since, in a pp-wave spacetime, the coordinate $u$
is a parameter that can label points on a geodesic curve (observer) and $u$
is identified (except for a constant $p$) with the worldsheet coordinate $\tau$, the change of shape from a circle to an ellipse
in the $xy$ plane
(before and after the pulse) is seen only in the $xyu$ embedding. In the
other embeddings, the string is a closed space curve (not circular),
as is visible in Figure 4.

\section{Conclusions}

\noindent Our main objective in this article was to show how a gravitational
wave pulse can affect the shape of string as it evolves to generate the
worldsheet. As mentioned in the Introduction, our motivation was to
provide a `stringy' realisation of the often used picture where memory is
linked to the `permanent change' in the shape of a ring of particles.
We achieve this goal by studying strings in pp-wave spacetimes where
the gravitational wave pulse is modeled by the free function available 
in such spacetimes. Studying strings in the presence of a square and
a sech-squared pulse we show that a circular string in the past
(before the arrival of the pulse) evolves into a deformed shape (ellipse, in the $xy$ plane) in the
asymptotic future. The satisfying fact is that we are able to demonstrate this
explicitly by solving all relevant equations analytically. \\
\noindent It may be argued that the shape of the string is a coordinate 
dependent feature and one should look for quantities which do not
depend on the choice of background coordinates. We have shown, towards the end
of our article, that there exist such quantities which may exhibit a permanent
change -- the behaviour of the metric determinant and the Ricci scalar of the
worldsheet metric being two such quantities. Both of them could possibly exhibit a `permanent
change', the details of which have been demonstrated clearly in the previous section. In addition, the different embeddings of the worldsheet in a three dimensional Euclidean background
provide a useful way of visualising the changes that occur after the
duration of the pulse. One may also consider finding the energy (target space) of the
string using the expression $P^0= -T_0 \int_0^{2\pi} \sqrt{-\gamma}\gamma^{\tau a} \left (\partial_a x^0 \right ) d\sigma$. Working through this calculation (not quoted here) we did find a difference between the expressions for `before' and `after' the pulse, which is also a signature of memory. \\
\noindent It is possible to work on this topic further and find more
examples. We have indeed found very similar effects for a triangular pulse
as well as a completely general pulse shape replacing the non-zero constant
in the square case. In all scenarios though we do find a `permanent change'.
One is tempted to ask therefore--what kind of pulse will show no
change in shape of a string? Another open question is related to the use
of a more general Kundt wave spacetime background. One can investigate strings
in such Kundt spacetimes and look for a similar memory effect \cite{ic2}, \cite{ic3} and its consequences. 

\section*{Acknowledgements}

\noindent AD thanks the Department of Physics, IIT Kharagpur, India for 
the opportunity to do a part of this work during his fourth year bachelors project
in 2023-24.

\end{document}